\documentclass[]{iopart}

\usepackage{hyperref}
\usepackage{cleveref}
\usepackage{iopams}  
\usepackage{cite}
\usepackage{graphicx}
\usepackage{epstopdf}

\begin{document}

\title[Accurate density measurement of a cold Rydberg gas]{Accurate density measurement of a cold Rydberg gas via non collisional two-body transitions}

\author{Anne Cournol, Jacques Robert, Pierre Pillet, Nicolas Vanhaecke}

\address{ Laboratoire Aim\'{e} Cotton, CNRS, B\^{a}t 505, Universit\'{e} Paris-Sud, 91405 Orsay, France}
\ead{anne.cournol@univ-paris13.fr}

\begin{abstract}
We experimentally demonstrate an original method to measure very accurately the density of a frozen Rydberg gas. 
It is based on the use of adiabatic transitions induced by the long-range dipole-dipole interaction in pairs of nearest neighbor Rydberg atoms by sweeping an electric field with time. 
The efficiency of this two-body process is experimentally tunable, depends strongly on the density of the gas and can be accurately calculated.
The analysis of this efficiency leads to an accurate determination of the Rydberg gas density, and to a calibration of the Rydberg detection.
Our method does not require any prior knowledge or estimation of the volume occupied by the Rydberg gas, or of the efficiency of the detection.

\end{abstract}


\maketitle

\section{Introduction}

Standard density measurements are indirect optical methods based on the interaction between an atomic or a molecular gas with a resonant light \cite{Wu:86}. Based on one-body interaction processes, the density is usually measured by determining the number of particles within an interaction volume and require the knowledge of this interaction volume.
In this article we report on an original experimental method which gives access to the nearest-neighbor distribution in a frozen gas of Rydberg atoms or molecules. 
We illustrate it with a very accurate, direct determination of the density of a Rydberg atomic sodium gas. 
It should be stressed that this determination does not require the knowledge of the volume occupied by the Rydberg gas and gives a direct access to the gas density.
Combined with the knowledge of this volume, our method would provide an accurate calibration of the detector itself.
The method could also be a tool to probe anisotropic nearest-neighbor distributions, for instance Rydberg gases in the blockade, antiblockade, and broken blockade regimes \cite{PhysRevLett.104.013001,PhysRevLett.102.013004}.
The method presented can be extended to many other atomic and molecular species, to more and less dense gases, both in beams and trapped gases.

The exaggerated electric dipole moments of Rydberg atoms or molecules are responsible for their strong dipole-dipole interactions \cite{Gallagher2008,Pillet2016}. These long-range interactions in cold and ultra-cold atomic samples have constantly received considerable attention since the dipole blockade was theoretically proposed to entangle neutral atoms with many applications in quantum optics, quantum computation and quantum simulation \cite{jaksch2000, lukin2001}. These systems have been experimentally investigated in number of context, leading to the observation of entangled many-body states \cite{vogt2006,heidemann2008,gaetan2009,urban2009}. The experimental investigations from few to many-body interactions include Rydberg atom counting statistics \cite{cubelliebisch2005,PhysRevLett.109.053002,PhysRevLett.113.023006}, Rabi oscillations \cite{NaturePhysics1183,PhysRevLett.100.113003}, echo experiments \cite{PhysRevLett.100.013002,youngeraithelnjp2008}, atom-pair interferometry \cite{PhysRevX.2.031011}, coherent dipolar coupling \cite{PhysRevX.2.031011,NaturePhysics10914}, three-body effect \cite{PhysRevLett.112.183002,Faoro:NatureComm10.1038} and microwave spectroscopy \cite{PhysRevLett.115.013001}. Recently, it has been shown that these strongly interacting many-body systems can lead to crystallization  and aggregation \cite{Schau1455,PhysRevLett.114.203002}.
More generally, many-body (including binary) processes are very much density dependent. Knowledge of spatial distribution and correlations is valuable to further study such atomic systems \cite{PhysRevA.72.063403,schausz2012,PhysRevA.87.053414,PhysRevA.88.061406}, and more complex systems as Rydberg molecules \cite{Gaj:NatureComm,Niederprum2016}. As such, the nearest-neighbor distribution in a Rydberg gas would give deep insight into the two-body positional correlation, which is of great interest in the dipole blockade regime. 
Hence, an accurate determination of the density of a gas phase sample is of great importance. 
This task is rather difficult and often presents large uncertainties which possibly hamper the achievable accuracy on other physical quantities. For instance, the evolution of Rydberg atomic and molecular gases towards plasmas depends strongly on the gas density \cite{PhysRevA.70.042713,PhysRevLett.101.205005,PhysRevLett.110.045004,PhysRevLett.110.253003,PhysRevA.89.022701}. Also, the determination of absolute cross sections of molecular collisions is a notoriously delicate task, since it requires an accurate knowledge of gas densities \cite{book:ramsey}.  

\begin{figure}
\centerline{\resizebox{0.7\linewidth}{!}{
\includegraphics{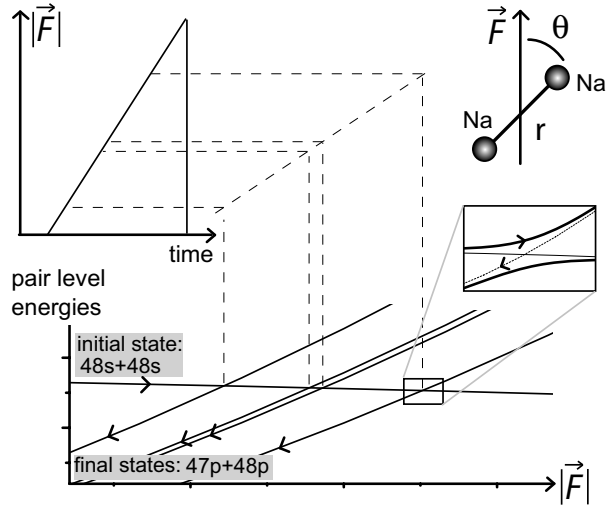}
}}
\caption{Top left: schematic representation of the linearly swept bias electric field which crosses the four 48{\it s}+48{\it s} - 47{\it p}+48{\it p} energy resonances (bottom).
The arrows symbolize successful transitions of pairs. 
Inset: schematic removal of an energy degeneracy by the dipole-dipole interaction.
Top right: parameters $r$ and $\theta$ describing a pair of neighboring Rydberg sodium atoms in the homogeneous electric field $\vec{F}$.
}
\label{fig:1}
\end{figure}


Our density measurement is a practical application of Landau-Zener transition through a F\"orster resonance. Those resonances arise from long-range dipole-dipole interaction between Rydberg particles tuned by an electric field. Dipole-dipole interaction scales with $r^{-3}$, where $r$ is the interparticle distance.
By varying in time a bias electric field in a frozen sample of Rydberg atoms, which interact through the dipole-dipole interaction, one can induce transitions in the internal state of pairs of neighboring atoms, several microns apart from each other \cite{PhysRevLett.104.133003}.
The induced transition in such pairs is a non collisional process, since the relative motion of the atoms is negligible on the timescale of the transition. 
Moreover, this process happens at very large distances with respect to both the size of the atoms and the de Broglie wavelength, such that the external degrees of freedom of the atoms can be treated classically.
Therefore, not only does this process not depend on any short-range interaction between the atoms,
but the efficiency of the transition, averaged over the Rydberg sample, depends only on the nearest-neighbor distribution. 
This efficiency can be analytically calculated as a function of the nearest-neighbor distribution and the sweeping rate of the bias electric field. 
In the present work, we operate in a spatially homogeneous sample, in which case the nearest-neighbor distribution depends on a single parameter: the gas density.
We characterize experimentally the amount of induced transitions for a number of values of the electric field sweeping rate, which leads us to a very accurate determination of the Rydberg gas density.

\section{The experimental setup}\label{sec:experiment}
The experimental setup consists of an atomic beam machine \cite{PhysRevLett.104.133003}.
A pulsed, seeded supersonic beam of sodium atoms is produced by expanding helium in vacuum, using laser-induced ablation of solid sodium.
The beam passes through a 2\,mm-diameter skimmer 20\,mm downstream from the nozzle, yielding a beam of ground-state sodium atoms at a translational temperature of about 1\,K.
The atoms are laser-excited to the 48\,$s$ state with a two-stage 3\,$s_{1/2}$ $\rightarrow$ 4\,$p_{3/2}$ $\rightarrow$ 48\,$s$ excitation, driven by a pulsed UV laser and a cw IR laser, respectively.  
The Rydberg excitation is done with unfocused laser beams that are baffled by 1\,mm-diameter diaphragms, which ensure that the Rydberg sample is uniformly distributed within the excitation volume, and that this excitation volume is constant.
After a time delay of 600\,ns following the Rydberg excitation, a small electric field is ramped up linearly with time, crossing the 48{\it s}\,0+48{\it s}\,0 - 47{\it p\,m}+48{\it p\,m$^\prime$} accidental energy degeneracies (so-called F\"{o}rster resonances \cite{vogt2006}), as described in Fig.\ref{fig:1}.
In the above notation, the third quantum number describing the state of each atom is the magnetic quantum number quantized on the electric field axis. 
There are several energy degeneracies, due to the non degenerate structure of both $p$ states in an electric field. 
These degeneracies are lifted by the dipole-dipole interaction, as shown in the inset of Fig.\ref{fig:1}.
The linear ramp is provided by a voltage function synthesizer with various sweeping rates $F'$, ranging from 0.6V/cm/$\mu$s to 7.8V/cm/$\mu$s.
The switching off of the electric field is done within 50\,ns.
This time-dependent electric field gives rise to transitions in the Rydberg pair, as symbolized by the arrows in Fig.\ref{fig:1}.

The atoms are subsequently ionized by a partially selective pulsed electric field. 
To do so, an electric field is ramped up through the various ionization thresholds, the ions produced are accelerated by the electric field and detected by micro-channel plates (MCPs). 
The arrival time of each ion on the MCPs reflects the state of the atom it originates from.
Atoms in the 48\,$p$ state are ionized at a lower electric field, i.e. earlier, than atoms in the 47\,$p$ or 48\,$s$ states. 
Gating the MCP signal allows us to produce a first integrated signal, $S^{48p}$, of ions originating from atoms in the 48\,$p$ state in a first gate and a second integrated signal $S^R$ of ions originating from the others states in a second gate. 
The total ion signal, $S^{T}$, is the sum of the first and second signals.
It is essential that the various Rydberg atoms are detected with the same efficiency. 
This is ensured by the rapidly rising ionization field, which reaches a final value that is much higher that the ionization threshold of the involved Rydberg states.

For a set of values of the sweeping rate $F'$, we record the integrated ion signal $S^{48p}$ arising from the 48\,$p$ atoms and the total ion signal $S^{T}$ arising from all Rydberg atoms.
The value of $F'$ reflects in the adiabaticity of the atom pair crossings, hence in the fraction of atoms converted from 48\,$s$ into 48\,$p$, which in turn reflects in the ratio between the ion signals $S^{48p}$ and $S^{T}$.



\section{Theoretical model} \label{theory}
\begin{figure}
	\centerline{ \resizebox{0.7\linewidth}{!}{
			\includegraphics{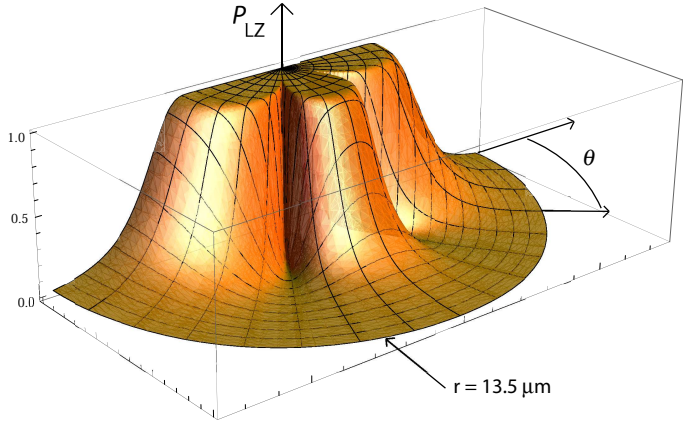}
		}}
		\caption{Polar plot of the adiabatic transition probability given by equation \eref{eq:PLZ} as a function of $r$ and $\theta$. The transition considered here is $n${\it s}\,0+$n${\it s}\,0 - ($n$-1){\it p}\,0+$n${\it p}\,0. Using an electric slew rate of $F'$=$1.0$\,V/cm/$\mu$s and the principal quantum number $n$=48 gives the spatial scale, indicated by the arrow pointing at the interatomic distance of 13.5\,$\mu$m.
		}
		\label{fig:2}
	\end{figure}

In this section, we make explicit the theoretical model applied to this system under two assumptions. 
Firstly if several crossings are involved and given the experimental slew rate of the bias electric field, one can prove that each crossing can be treated independently and successively by estimating the transition time $\tau_{LZ}$ for one isolated adiabatic crossing in a two atom-pair states system \cite{PhysRevA.59.988}. Considering our typical experimental conditions, we estimate $\tau_{LZ} \approx 5$ ns, corresponding for the higher electric field slew rate to $\Delta F \approx 70$ mV/cm which is smaller than the splitting between two adjacent crossings. 
Secondly due to the time dependence range of the bias electric field, each energy degeneracy can clearly give rise only once to adiabatic transitions in a pair of neighboring atoms, i.e., only when the field is ramped up (see Fig.\ref{fig:1}).
Indeed, the switching off of the electric field is done abruptly enough to ensure that the state of all atom pairs is diabatically conserved.
Therefore, the series of crossings of Fig.\ref{fig:1} is well suited for a treatment in the frame of a Landau-Zener model  \cite{landau1932,zener1932} by treating once each crossing independently. In this section, we first develop the theoretical model for a single adiabatic crossing. We then link it to the treatment of several independent crossings.  
Since the interaction that lifts the energy degeneracy is the dipole-dipole interaction, the probability for an adiabatic transition for a single crossing in a given pair of atoms initially prepared in the $ns$ state reads
\begin{equation}\label{eq:PLZ}
  P_{\textrm{\footnotesize{LZ}}}(F',\,r,\,\theta)=1 - \exp \left[- \left( \frac{r}{r_0(F',\theta)}\right)^{-6} \right]
\end{equation}
where $r$ is the interatomic distance, $\theta$ is the polar angle of the interatomic direction with respect to the electric field direction (see Fig.\ref{fig:1} top right) and 
\begin{eqnarray*}
r_0 (F',\,\theta) 
& = & 
\left[ 
\frac{2 \pi \big| \langle ns \| \mu \| np \rangle \langle ns \| \mu \| (n-1)p \rangle f(\theta)^3
 \big| ^{2}}
{(4\pi \epsilon_0)^{2} \hbar\, | \Delta\mu | \, | F' |}
\right]^{1/6} \\
& = & 
r_0 (F') f(\theta),
\end{eqnarray*}
where $ \Delta\mu $ stands for the effective dipole moment difference between diabatic pair states, $F'$ is the time derivative of the electric field experienced by the atom pair when the electric field is ramped up (see Fig.\ref{fig:1}).
The factor $ \langle ns \| \mu \| n'p \rangle $ denotes the reduced matrix element of the electric dipole moment operator $\mu$. The term $f(\theta)$ encompasses the angular dependence of the dipole-dipole interaction, and depends on the magnetic quantum numbers of the diabatic pair states, which are quantized on the local applied electric field axis. 
Fig.\ref{fig:2} shows a polar plot of $P_{\textrm{\footnotesize{LZ}}}$ for a transition through the resonance $ns\,0$+$ns\,0$-$(n-1)p\,0$+$np\,0$, with $n$=48, for which $f(\theta)^3=\cos^2\theta-1/3$.

Let us denote $\xi(r,\theta)$ the nearest-neighbor distribution of the Rydberg gas, which we consider translationaly invariant but possibly anisotropic.
For a given time derivative of the electric field $F'$, the expected value of the transition probability $\left< ns,ns \rightarrow np,(n-1)p \right>_{F'}$ reads
\begin{eqnarray}\label{ExpValTrans}
\int_{0}^{2\pi}
\int_{0}^{\pi}
\int_{0}^\infty \xi(r,\theta) P_{\textrm{\footnotesize{LZ}}}(F',r,\theta)
\,d\phi\,\sin(\theta) d\theta\, r^2 dr .
\end{eqnarray}
If the Rydberg gas is spatially uniformly distributed with a density $\eta$, the nearest-neighbor distribution is given by the Erlang probability density function \cite{book:statisticaldistributions2000}
\begin{eqnarray*}
\xi_\eta(r,\theta)=\eta \, \exp \big[ -\frac{4\pi}{3}\eta r^3 \big]
,
\end{eqnarray*}
and the expected value of the transition $\left<ns,ns \rightarrow (n-1)p,np\right>_{\eta,\,F'}$ also depends parametrically on the density $\eta$. By integrating over the variable $r$ the expected value can be expressed with the help of the special Meijer $G$ function \cite{book:handbookInt2000}. For one crossing, $\left<ns,ns \rightarrow (n-1)p,np\right>_{\eta,\,F'}$ reads (see appendix \ref{app:singlecrossing})
\begin{equation}
\begin{array}{l}
\label{eq:meijer}
1-
\frac{1}{2\sqrt{\pi}}
\int_{0}^{\pi}
\sin(\theta) d\theta\,
G^{3\,0}_{0\,3} 
\left( \frac{1}{4} 
\left( \frac{r_0(F')}{a_\eta} \right)^6
f(\theta)^6
\Big| ^{\;\;\,-}_{0\,\frac{1}{2}\,1} 
\right) \quad ,
\end{array}
\end{equation}
where $a_\eta$ is the Wigner-Seitz radius, defined by $\frac{4\pi}{3}a_\eta^3=\eta^{-1}$.\\

Eq.\ref{eq:meijer} is a convenient expression in the case of several independent crossings. Indeed, summing individual probabilities leads to simply substituting $f(\theta)^6$ by $\sum_i f_i(\theta)^6$ where $i$ denotes the crossing (see appendix \ref{app:multicrossing}).
This provides an analytic expression for the production of $(n-1)p + np$ pairs starting from $ns$ atoms {\it via} adiabatic transitions in pairs, as a function of the Rydberg gas density $\eta$.

In the experimental volume $V$ occupied by the Rydberg gas which contains initially $\eta V$ atoms in the $ns$ state, the number of $np$ atoms produced by Landau-Zener transitions for a given density $\eta$ and a given value of $F'$ reads 
\begin{eqnarray*}
\frac{1}{2}\, \eta \, V
\left< ns,ns \rightarrow (n-1)p,np \right>_{\eta,\,F'}.
\end{eqnarray*}
The factor 1/2 accounts for the fact that one transition from a pair of atoms in the $ns$ state gives rise to one atom in the $np$ state.

The total number of Rydberg atoms is clearly proportional to the Rydberg density, with a proportionality factor which amounts to the volume $V$. 
In addition, the total integrated ion signal $S^T$ depends on the total number of Rydberg atoms through the efficiency of the detection, which accounts for the ion collection and the efficiency of the MCP detector. 
A key assumption of the model is that the volume $V$ is constant (appendix \ref{app:fluctuations} provides a justification a posteriori of this important assumption).
This allows us to write $S^T=h(\eta)$, where $h$ is a monotonically increasing function of $\eta$.
As mentioned above, thanks to the detection of the Rydberg states based on field ionization, all the considered Rydberg states, {\it i.e.}, $ns$, $np$, and $(n-1)p$ are detected with the same efficiency.
Therefore, not only does the total number of Rydberg atoms read $\eta\,V = g(S^T)V$, where $g$ is the inverse function of the above defined function $h$, but the number of $np$ atoms reads $g(S^{np})V$, where $S^{np}$ is the ion signal originating from $np$ Rydberg atoms.
Finally, the ion signal $S^{np}$ can be analytically linked to the total ion signal $S^T$:
\begin{equation} \label{eq:fctfit}
S^{np}=h\left( \frac{1}{2} g\left( S^T \right)  
\left< ns,ns \rightarrow (n-1)p,np \right>_{g(S^T),F'} \right)
\quad ,
\end{equation}
where $\left< ns,ns \rightarrow (n-1)p,np \right>_{g(S^T),F'}$ is given by Eq.\ref{eq:meijer}.
Under an additional constrain that the function $h$ (or alternatively the function $g$) belongs to a given family of functions, for instance a polynomial function of a given degree, the function $h$ and its inverse function $g$ can be fully characterized by comparing the above model with the experimental data. This is the topic of the following section.

\section{Results and discussion}\label{sec:results}
In Fig.\,\ref{fig:3} we show more than 75000 single-shot experimental data points (dots) representing the signal $S^{np}$ as a function of both $S^T$ and $1/ F'$, where $n$=48.
Measurements without the swept electric field have been done to quantify the effect of black-body radiation on $ns$ atoms, which may drive them into the $np$ state; all data presented in Fig.\,\ref{fig:3} have been corrected for this effect.
Experimentally, the range of $S^T$ is spanned thanks to shot-to-shot fluctuations due to both the variations of the beam seeding efficiency and the frequency jitter of the pulsed UV laser. 
We fit these experimental data with a two-variable ($S^T$, $F'$) 
least-square adjustment using Eq.\ref{eq:fctfit}, as expained in the following sections.

\subsection{Linear model}\label{subsec:linear}
In a first approach we consider that $S^T$ depends linearly on the density $\eta$.
This approach seems very reasonable, since in our experiment the MCP detector used for the detection of the Rydberg atoms is operated at low gain, where a linear regime is expected.
In the frame of this linear model, $\eta$ reads $g(S^T)=g_0\,S^T$, where $g_0$ is the single unknown scalar parameter of the model. 
The least-square adjustment converges to the best fit parameter $\hat{g}_0=4.150\times 10^{15}\,$cm$^{-3}$/(V\,s) with a standard deviation of $4\times 10^{12}\,$cm$^{-3}$/(V\,s).

In addition to the above statistical noise arising from the fit itself, an additional uncertainty arises from the fact that the Rydberg sample is not infinite. 
To take this effect into account, the expected value in Eq.\ref{eq:fctfit} should be replaced by a sample mean over the Rydberg ensemble, which is estimated as follows.
The outcome of the transition in each pair of atoms follows a binomial distribution with a probability $P_{\textrm{\footnotesize{LZ}}}(r)$. The variance of this discrete process is at most 1/4. Thus, relying on the central limit theorem, given a number of atoms of typically $\approx10^4$, this leads to an accuracy on the determination of the averaged value of the density over signal ratio $g$ better than 4\%.
Typically, one experimental shot of $S^T=10$\,nVs corresponds to $\eta=4.2 \pm 0.2 \times 10^7 \mathrm{cm}^{-3}$. This agrees with previous laser-induced-fluorescence measurements which we had done on the ground state sodium atoms in our beam \cite{PhysRevLett.104.133003}.

\subsection{Quadratic model}\label{subsec:quadratic}
In a second, refined approach we consider that $S^T$ involves in addition a quadratic dependence on the density $\eta$. Therefore we write now $g(S^T)=g_0 S^T + g_1 S^{T\,2}$, thereby involving two unknown scalar parameters $g_0$ and $g_1$.
The least-square adjustment converges to the best fit parameters $\hat{g}_0=3.039\times 10^{15}\,$cm$^{-3}$/(V\,s) and $\hat{g}_1=2.80\times 10^{10}\,$cm$^{-3}$/(V\,s)$^2$ with relative standard deviations of $10^{-3}$ and $10^{-2}$, respectively.
As expected, the quadratic parameter is found to be a correction to the linear term, of positive sign, thereby reflecting a slight saturation effect of the MCP detector (i.e. a saturation in the corresponding inverse function $h(\eta)$).  
A compared analysis of variance of the linear and quadratic models allows us to conclude that the quadratic model is significantly better. In other words it is very unlikely that the better agreement with the quadratic model would solely be due to statistical randomness.
Of course, an uncertainty arising from the finiteness of the Rydberg sample also arises in this quadratic model, and the remarks above apply also to the outcome of the fit with the quadratic model.

\begin{figure}
\centerline{ \resizebox{0.7\linewidth}{!}{
		\includegraphics{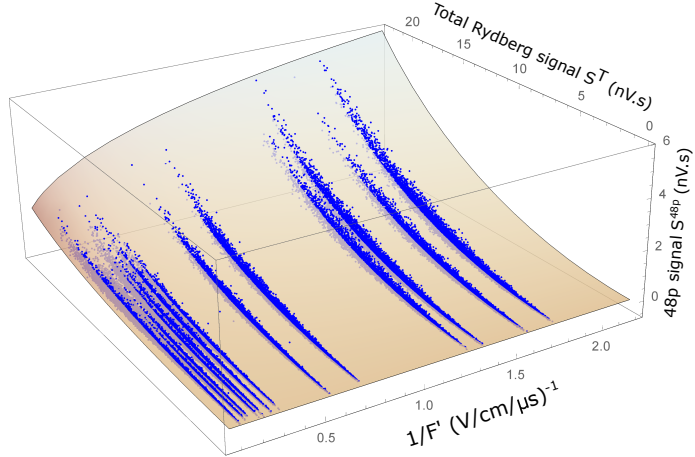}
	}}
    \caption{Dots: single-shot experimental data points of the $n\,p$ signal, $S^p$, where $n$=48, as a function of the total integrated ion signal arising from the Rydberg gas $S^T$ and the inverse of the time-derivative of the applied electric field.
Surface: best fit function given by Eq.\ref{eq:fctfit} with the best fit parameters of the quadratic model (see text section \ref{subsec:quadratic}). 
}
    \label{fig:3}
\end{figure}

In Fig.\,\ref{fig:3} we also show a surface plot of the best fit function given by Eq.\ref{eq:fctfit} with the above best fit parameters.
Experimental data points are visible by virtue of the transparency of the surface if they are located below the surface.
Fig.\,\ref{fig:3} shows a remarkably good agreement between the fitted model and the experimental data, which manifests itself in trend-free residuals.
The reduced chi-square value obtained amounts to 0.020\,(nVs)$^2$, i.e., an estimation of the experimental noise is 0.14\,nVs \cite{Amsler20081}, which is in quantitative agreement with the scattering of the experimental signal $S^{np}$ shown in Fig.\,\ref{fig:3}.


\subsection{Conclusion}
In the frame of a given model, such as one of the models discussed above, the best fit parameters characterizing the function $g$ (or $h$) provide a conversion function between the density of the Rydberg gas and the recorded signal.
We have presented the fitted conversion function in the frame of a linear and a quadratic model, but the extension to higher order model is of course straightforward. Standard statistics offer many techniques to conclude on most adequate description of the conversion between the density of the Rydberg gas and the recorded signal \cite{book:gelman}. 

In fact, the relationship between the density of the Rydberg gas and the recorded signal certainly fluctuates from shot to shot, due to the fluctuations of both the detection efficiency and the volume in which the atoms have been prepared in the initial Rydberg state.
Therefore the fitted conversion function between the density of the Rydberg gas and the recorded signal $S^T$ should be understood as mean conversion function. 
We show in appendix \ref{app:fluctuations} that it is possible to estimate the fluctuations of the conversion function, based on an analysis of the scattering of the experimental data of Fig.\,\ref{fig:3} and on a differentiation of Eq.\ref{eq:fctfit}.
In addition the same analysis of appendix \ref{app:fluctuations} shows a posteriori that the requirement of a constant volume $V$ is indeed fulfilled under our experimental conditions.

It should be insisted on the fact that the presented density measurement requires knowledge neither of the volume occupied by the Rydberg gas, nor of the efficiency of the detector, and it provides a direct calibration of its density.
Together with a precise knowledge of this volume, such a density measurement constitutes an accurate calibration of the detection chain, including ion optics and detector efficiency.
Since ion optics can be precisely numerically simulated, the presented method can provide a very accurate calibration of an ion detector, as well as its response statistics, in the keV energy range.

\section{Conclusion}
Many Rydberg pair systems present avoided crossings in an electric field, which makes the present technique applicable to a broad range of systems.
Our method is applicable in Rydberg gases less dense than in this work, since one can induce dipole-dipole induced transitions in pairs at even longer distances than we do: this can be done either by exciting higher Rydberg states which exhibit stronger dipole-dipole interactions, or by sweeping the electric field even slower, which can be done as long as the atoms are still frozen during the duration of the adiabatic transition.
Higher density gases can also be probed with our method, as long as no three or more-body effects come into play. 
This shall provide a very valuable tool for measuring accurately the density in experiments dealing with cold Rydberg atomic or molecular samples leading to a plasma phase \cite{PhysRevLett.101.205005}.
In this context, as well as in others where non linear density-dependent processes are involved, the amplitude of the shot-to-shot scattering of the gas density provided by the present method would be of great value in the analysis of the processes.
Moreover, the breakdown of the model presented in this article at higher gas densities would be a signature of three-body processes.

The presented method could also give access to the nearest-neighbor distribution in ultracold samples of Rydberg in the dipole blockade regime \cite{gaetan2009,urban2009}.
As a matter of fact, it could provide a direct proof of the isolation of the Rydberg atoms created in the dipole blockade regime.
The method can easily be adapted to probe an anisotropic nearest-neighbor distribution. 
Indeed, by isolating a single avoided crossing showing an anisotropy of the transition probability (as shown in Fig.\ref{fig:2}), one can explore the nearest-neighbor distribution as a function of the direction of the bias electric field.
This could be of great interest in Rydberg gases in the dipole blockade, antiblockade, and broken blockade regimes \cite{PhysRevLett.104.013001,PhysRevLett.102.013004}, which are induced by the anisotropic dipole-dipole interaction.

\section*{Acknowledgments}
This work has been financed by the ANR grant CORYMOL (No. NT05-2 41884) and by the ``Institut Francilien de Recherche sur les Atomes Froids'' (IFRAF).
The authors thank D.~Comparat and E.~Brion for numerous fruitful discussions.
The authors are very grateful to F.~Merkt for his friendly support and constant interest in this work.

\section{Appendices}


\subsection{Transition probability for a single crossing}\label{app:singlecrossing}
In this appendix we derive the analytical expression of the expected value of the transition probability for a single crossing.
Eq.\ref{ExpValTrans} represents the average over the atomic sample of the adiabatic transition probability in the limit where the number of atoms is large with the help of the Erlang probability density function. The integral brings up the sum of two terms:
\begin{eqnarray}
2\pi\eta
\int_{0}^{\pi}
\int_{0}^\infty
\exp \left[- \left( \frac{r}{a_{\eta}}\right)^{3} \right] 
\sin(\theta) d\theta\, r^2 dr, \label{firstterm}
\\
2\pi\eta
\int_{0}^{\pi}
\int_{0}^\infty 
\exp \left[- \left( \frac{r}{a_{\eta}}\right)^{3} + \left(\frac{r}{r_0 (F') f(\theta)}\right)^{-6}  \right]
\sin(\theta) d\theta\, r^2 dr . \label{secondterm}
\end{eqnarray}
\\
The term (\ref{firstterm}) corresponds to the integration over the all space of the Erlang probability density function and is equal to 1.
The integration over $r$ in term (\ref{secondterm}) can be express analytically with the help of the special Meijer $G$ function \cite{book:handbookInt2000}:

\begin{eqnarray*}
\hspace*{-2.5cm}
\int_{0}^\infty 
\exp \left[- \left( \frac{r}{a_{\eta}}\right)^{3} + \left(\frac{r}{r_0 (F') f(\theta)}\right)^{-6}  \right]
 r^2 dr
=
\frac{1}{6}\sqrt{\frac{r_0 (F')^6 f(\theta)^6}{\pi}}
G^{3\,0}_{0\,3} 
\left( \frac{1}{4} 
\left( \frac{r_0(F')}{a_\eta} \right)^6
f(\theta)^6
\Big| ^{\;\;\;\;\;-}_{-\frac{1}{2}\,0\,\frac{1}{2}} 
\right).
\end{eqnarray*}
\\
Using the property \cite{book:handbookInt2000}:$
\hspace*{1.5cm}
z^{\,\frac{1}{2}}\;
G^{3\,0}_{0\,3} 
\left( z
\Big| ^{\;\;\;\;\;-}_{-\frac{1}{2}\,0\,\frac{1}{2}} 
\right)
=
G^{3\,0}_{0\,3} 
\left(z
\Big| ^{\;\;\;-}_{0\,\frac{1}{2}\,1} 
\right)
$,
\\
term (\ref{secondterm}) finally reads

\begin{eqnarray*}
\hspace*{0cm}
\frac{1}{2\sqrt{\pi}}
\int_{0}^{\pi}
\sin(\theta) d\theta\,
G^{3\,0}_{0\,3} 
\left( \frac{1}{4} 
\left( \frac{r_0(F')}{a_\eta} \right)^6
f(\theta)^6
\Big| ^{\;\;\,-}_{0\,\frac{1}{2}\,1} 
\right).
\end{eqnarray*}
\\
This compact analytical expression is used for the least-square adjustments of section \ref{sec:results}.

\subsection{Transition probability for multiple crossings}\label{app:multicrossing}
We now consider a series of $N$ independent crossings in the frame of the Landau-Zener model. The total probability for an adiabatic transition through $N$ independent crossings reads

\begin{equation*}
  P_{\textrm{\footnotesize{LZ,total}}}(F',\,r,\,\theta)=1 - \overline{P}_{\textrm{\footnotesize{LZ,total}}}(F',\,r,\,\theta).
\end{equation*}
We define $\overline{P}_{\textrm{\footnotesize{LZ}},i}(F',\,r,\,\theta)$ the probability for a diabatic transition through the crossing $i$. The total diabatic probability $\overline{P}_{\textrm{\footnotesize{LZ,total}}}(F',\,r,\,\theta)$ reads

\begin{eqnarray*}
\overline{P}_{\textrm{\footnotesize{LZ,total}}}(F',\,r,\,\theta)
& = & 
{\displaystyle \prod_{i=1}^{N} \overline{P}_{\textrm{\footnotesize{LZ}},i}(F',\,r,\,\theta)}  \\
& = & 
{\displaystyle \prod_{i=1}^{N} \exp \left[- \left( \frac{r}{r_0(F')f_i(\theta)}\right)^{-6} \right]} \\
& = &
\exp \left[- \left(\frac{r}{r_0(F')}\right)^{-6} \displaystyle \sum_{i=1}^{N}f^{\;6}_i(\theta) \right]
\end{eqnarray*}
Considering the magnetic sub-levels $\{ ns0+ns0,(n-1)pm+npm' \}$ gives rise to 4 resonances (see Fig.\ref{fig:1}). It is worth noting that $ \sum_{i=1}^{4} f^{\;6}_i(\theta) = 2/3$, the total probability for an adiabatic transition through the multiplicity no longer depends on $\theta$. The probed volume around a Rydberg atom for a nearest neighbor corresponds to a "sphere" of a soft radius $\frac{2}{3}r_0(F')$.

\subsection{Fluctuations of the conversion function}\label{app:fluctuations}
In this appendix we show an estimation of the fluctuations of the conversion between the density of the Rydberg gas and the recorded signals.

In the following we derive the estimation of these fluctuations in the frame of the linear model of section \ref{subsec:linear}.
The experimental signal ions are integrated by two gates: the first gate records the $np$ ion signal, $S^{np}$, while the second gate records the rest of the ion signal, $S^R$. The total ion signal amounts to $S^T=S^{np}+S^R$. 
Let us assume that the noise induced on both of these boxes is independent and uncorrelated. This is reasonable since the integrated signals are 
essentially due to fluctuations in the detection efficiency and possibly stray ions. 
Therefore, the shot-to-shot conversion might not be the same for the $np$ signal and for the rest of the signal, and we introduce two different conversion factors, $g_{np}$ and $g_R$, with obvious notations.
Consequently, the $np$ signal reads (in the approximation of an infinite sample):
\begin{equation}
S^{np}=\frac{1}{2}
\frac{g_R S^R+g_{np} S^{np}}{g_{np}}
\left< ns\,ns \rightarrow (n-1)p\,np \right>_{g_R S^R+g_{np}S^{np},\,F'}
\quad.
\end{equation} 
Formally, since $S^T=S^{np}+S^R$, this can be written in an implicit form, which reads 
\begin{equation}\label{eq:implicitform}
\phi(
S^{np},S^R,g_{np},g_R
) = 0 \quad,
\end{equation}
where $S^{np}$, $S^R$, $g_{np}$, and $g_R$ are now all considered as independent variables (in addition $\phi$ depends parametrically on $F'$). 
This allows us to differentiate the above equation \ref{eq:implicitform} and, in a local linear approximation of $\phi$, derive the expression of the root mean square (rms) value of $g_{np}$, $\delta g_{np}$, as a function of the rms values of $S^{np}$ and $S^T$, $\delta S^{np}$ and $\delta S^T$, respectively.
Similarly, the rms value of $g_R$, $\delta g_R$, can be expressed as a function of $\delta S^{np}$ and $\delta S^T$.
This is done as follows.

 \begin{figure}
 \centerline{\resizebox{0.7\linewidth}{!}{
 \includegraphics{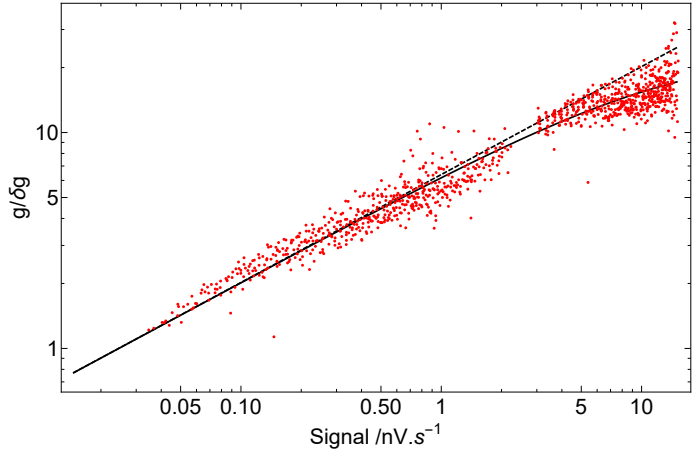}
 }}
     \caption{Signal to noise plot, displayed in a log-log plot.
 The dots are deduced from the experimental data by binning data along the $S^T$-axis to deduce the scattering of $S^{np}$, and compute $g/\delta g$ (see text). The left bunch of dots represents $g_{np}/\delta g_{np}$, and the right dots represent $g_R/\delta g_{R}$. The solid line is the best fit curve using a Polya distribution. The dashed line represents the signal to noise expected from a pure Poisson statistics. Deviation from the pure Poisson statistics is due to the fluctuations in the avalanche process in the micro-channel plates and to the fluctuations of the volume occupied by the Rydberg gas (see text).  
 }
     \label{fig:4}
 \end{figure}

Firstly, let us go back to the definition of $g$ as introduced in section \ref{theory}.
The total integrated ion signal $S^{np}$ depends on the total number of $np$ Rydberg atoms through the efficiency of the detection for the first gate, $\gamma_{np}$, which accounts for the Rydberg ionization, the ion collection and the efficiency of the MCP detector. 
In addition, the number of $np$ Rydberg atoms is proportional to the density of $np$ Rydberg atoms through the volume $V$. 
Therefore one has $g_{np} \gamma_{np} V = 1$.
Similarly one also has $g_R \gamma_R V = 1$, where $\gamma_R$ is the detection efficiency of the second gate.
The fluctuations of $g$ arise from the fluctuations of the dectection efficiency and the fluctuations of the volume, i.e.:
\begin{equation}
\left(\frac{\delta g}{g}\right)^2 
=
\left(\frac{\delta \gamma}{\gamma}\right)^2 
+
\left(\frac{\delta V}{V}\right)^2 
\quad ,
\end{equation}
where ($g$, $\delta g$, $\gamma$, $\delta \gamma$) stands for ($g_{np}$, $\delta g_{np}$, $\gamma_{np}$, $\delta \gamma_{np}$) or ($g_R$, $\delta g_R$, $\gamma_R$, $\delta \gamma_R$).

Secondly, we follow the approach described in \cite{book:PMTHandbook} in order to account for the statistics related to the production of secondary emission of electrons in the detector, where the signal-to-noise ratio is modelled by Polya distributions \cite{Prescott1966173}.
Therefore we consider that the fluctuations of the detection efficiencies $\gamma_{np}$ and $\gamma_R$ depend on the amount of detected signal in the corresponding gate, so that these dependencies follow the same Polya distribution. Namely,
\begin{equation}\label{eq:polyagamma}
\frac{\gamma}{\delta \gamma} 
=
\frac{\alpha S}{
\left(
\beta S^2+S
\right)^{1/2}
}
\end{equation}
where ($\gamma$, $\delta \gamma$, $S$) stands for ($\gamma_{np}$, $\delta \gamma_{np}$, $S^{np}$) or ($\gamma_R$, $\delta \gamma_R$, $S^R$), and where $\alpha$ and $\beta$ are unknown parameters which fully characterize the Polya distribution. 
Using the same Polya distribution for both gates is very reasonable, since both signals come from the same detector working at low gain, such that there is no significant dead time effect in the detection.
The relative fluctuation of the volume, however, is independent of the signal and taken as a constant $\Gamma$.
It follows that 
\begin{equation}\label{eq:polya}
\frac{g}{\delta g} 
=
\frac{\alpha S}{
\left(
\left(\beta+\alpha^2\Gamma^2 \right) S^2+S
\right)^{1/2}
}
\end{equation}
where ($g$, $\delta g$, $S$) stands for ($g_{np}$, $\delta g_{np}$, $S^{np}$) or ($g_R$, $\delta g_R$, $S^R$).

Thirdly, we insert the expression \ref{eq:polya} of the fluctuations of the detection efficiency in the differentiation of equation \ref{eq:implicitform}. This gives an expression for the rms of the random variable $g_R$:
\begin{eqnarray*}
\delta g_R & = & 
\left[
\left( \frac{\partial \phi}{\partial S^{np}} \right)^2
\delta S^{np \,2}
-
\left( \frac{\partial \phi}{\partial S^{T}} \right)^2
\delta S^{T \,2}
\right]^{1/2} \\
& & 
\quad / \, \left[
\left( \frac{\partial \phi}{\partial g_R} \right)^2
+
\left( \frac{\partial \phi}{\partial g_{np}} \right)^2
\frac{\beta +\alpha^2\Gamma^2 + 1/S^{np}}{\beta +\alpha^2\Gamma^2 + 1/S^{R}}
\right]^{1/2} 
\quad ,
\end{eqnarray*}
where $\delta S^{np}$ depends now on $S^T$, as a consequence of the scattering analysis using bins along the $S^T$ axis.
A similar expression can be deduced for $\delta g_{np}$: 
\begin{eqnarray*}
\delta g_{np} & = & 
\left[
\left( \frac{\partial \phi}{\partial S^{np}} \right)^2
\delta S^{np \,2}
-
\left( \frac{\partial \phi}{\partial S^{T}} \right)^2
\delta S^{T \,2}
\right]^{1/2} \\
& & 
\quad / \, 
\left[
\left( \frac{\partial \phi}{\partial g_R} \right)^2
\frac{\beta +\alpha^2\Gamma^2 + 1/S^{R}}{\beta +\alpha^2\Gamma^2 + 1/S^{np}}
+
\left( \frac{\partial \phi}{\partial g_{np}} \right)^2
\right]^{1/2}
\quad .
\end{eqnarray*}

Finally, the experimental scattering of the data points is analyzed by binning the data along the $S^T$ variable: for each bin of well-defined $S^T$ value, the scattering of the $S^{np}$ values is deduced.  From this scattering and the above expressions the signal to noise $g_{np}/\delta g_{np}$ and $g_R/\delta g_{R}$ are evaluated and fitted together in the same least square adjustment using the expression of the Polya distribution given by equation \ref{eq:polya}, where the unknown parameters of the fit are $\alpha$, $\beta$ and $\Gamma$. 
Since the unknown parameters of the fit are used to prepare the data to be fitted, $g_{np}/\delta g_{np}$ and $g_R/\delta g_{R}$, a single adjustment cannot be used. Instead, a series of iterative adjustments is used, with an initial set of parameters corresponding ot a pure Poisson distribution and no fluctuation of the volume (i.e. $\beta$=0 and $\Gamma$=0). 
In addition, the parameters $\beta$ and $\Gamma$ always appear linked in the form of $\beta+\alpha^2\Gamma^2$, so that $\beta$ and $\Gamma$ cannot be both determined. 

We first assume no fluctuation of the volume.
Figure \ref{fig:4} shows the results of the series of adjustments, yielding the best fit parameters $\hat{\alpha}$=$6.4\,(nVs)^{-1/2}$ and $\hat{\beta}$=$0.072\,(nVs)^{-1}$.
The signal to noise is dominated by the emission of primary electrons in the MCPs, which alone would give a Poisson statistics (corresponding to $\beta$=0), represented by the dashed line in Fig.\ref{fig:4}.
The parameter $\beta$ reflects a deviation from the Poisson statistics, which is due to the fluctuations in the avalanche process taking place in the MPCs \cite{Prescott1966173}.

If we now assume that the volume may fluctuate, the previous results shows that an upper limit of the relative fluctuations of the volume is $\Gamma_{\textrm{\footnotesize{max}}}$=$\hat{\beta}^{1/2}/\hat{\alpha}$, which amounts to 4\%. This shows a posteriori that the experimental conditions detailed in section \ref{sec:experiment} indeed ensure a very constant volume.
Although the fluctuations of the volume cannot be discriminated from the fluctuations in the avalanche process taking place in the MPCs, the fluctuations of the conversion function $g$ are fully characterized by equation \ref{eq:polya} with $\alpha$=$6.4\,(nVs)^{-1/2}$ and $\beta+ \alpha^2\Gamma^2$=$0.072\,(nVs)^{-1}$.
 

Although we have presented here the estimation of the fluctuations of the conversion between Rydberg density and recorded ion signal in the frame of the linear model of section \ref{subsec:linear}, extension to higher order model (e.g. quadratic) is possible, although the corresponding derivations may become cumbersome.

\section*{References}

\bibliographystyle{unsrt}
\bibliography{bib2}

\begin{thebibliography}{10}

\bibitem{Wu:86}
Z.~Wu, M.~Kitano, W.~Happer, M.~Hou, and J.~Daniels.
\newblock Optical determination of alkali metal vapor number density using
  faraday rotation.
\newblock {\em Appl. Opt.}, 25(23):4483--4492, Dec 1986.

\bibitem{PhysRevLett.104.013001}
Thomas Amthor, Christian Giese, Christoph~S. Hofmann, and Matthias
  Weidem\"uller.
\newblock Evidence of antiblockade in an ultracold {Rydberg} gas.
\newblock {\em Phys. Rev. Lett.}, 104(1):013001, Jan 2010.

\bibitem{PhysRevLett.102.013004}
T.~Pohl and P.~R. Berman.
\newblock Breaking the dipole blockade: Nearly resonant dipole interactions in
  few-atom systems.
\newblock {\em Phys. Rev. Lett.}, 102(1):013004, Jan 2009.

\bibitem{Gallagher2008}
Thomas~F. Gallagher and Pierre Pillet.
\newblock Dipole-dipole interactions of {Rydberg} atoms.
\newblock volume~56 of {\em Advances In Atomic, Molecular, and Optical
  Physics}, pages 161 -- 218. Academic Press, 2008.

\bibitem{Pillet2016}
P~Pillet and T~F Gallagher.
\newblock {Rydberg atom interactions from 300 $\mu$K to 300 K}.
\newblock {\em Journal of Physics B: Atomic, Molecular and Optical Physics},
  49(17):174003, 2016.

\bibitem{jaksch2000}
D.~Jaksch, J.~I. Cirac, P.~Zoller, S.~L. Rolston, R.~C\^ot\'e, and M.~D. Lukin.
\newblock Fast quantum gates for neutral atoms.
\newblock {\em Phys. Rev. Lett.}, 85(10):2208--2211, Sep 2000.

\bibitem{lukin2001}
M.~D. Lukin, M.~Fleischhauer, R.~Cote, L.~M. Duan, D.~Jaksch, J.~I. Cirac, and
  P.~Zoller.
\newblock Dipole blockade and quantum information processing in mesoscopic
  atomic ensembles.
\newblock {\em Phys. Rev. Lett.}, 87(3):037901, Jun 2001.

\bibitem{vogt2006}
Thibault Vogt, Matthieu Viteau, Jianming Zhao, Amodsen Chotia, Daniel Comparat,
  and Pierre Pillet.
\newblock Dipole blockade at {F}\"orster resonances in high resolution laser
  excitation of {Rydberg} states of cesium atoms.
\newblock {\em Phys. Rev. Lett.}, 97(8):083003, 2006.

\bibitem{heidemann2008}
Rolf Heidemann, Ulrich Raitzsch, Vera Bendkowsky, Bj\"{o}rn Butscher, Robert
  L\"{o}w, and Tilman Pfau.
\newblock {Rydberg} excitation of {Bose}-{Einstein} condensates.
\newblock {\em Phys. Rev. Lett.}, 100(3):033601, 2008.

\bibitem{gaetan2009}
Alpha Ga\"etan, Yevhen Miroshnychenko, Tatjana Wilk, Amodsen Chotia, Matthieu
  Viteau, Daniel Comparat, Pierre Pillet, Antoine Browaeys, and Philippe
  Grangier.
\newblock Observation of collective excitation of two individual atoms in the
  {Rydberg} blockade regime.
\newblock {\em Nat Phys}, 5(2):115--118, February 2009.

\bibitem{urban2009}
E.~Urban, T.~A. Johnson, T.~Henage, L.~Isenhower, D.~D. Yavuz, T.~G. Walker,
  and M.~Saffman.
\newblock Observation of {Rydberg} blockade between two atoms.
\newblock {\em Nat Phys}, 5(2):110--114, February 2009.

\bibitem{cubelliebisch2005}
T.~Cubel~Liebisch, A.~Reinhard, P.~R. Berman, and G.~Raithel.
\newblock Atom counting statistics in ensembles of interacting {Rydberg} atoms.
\newblock {\em Phys. Rev. Lett.}, 95(25):253002, Dec 2005.

\bibitem{PhysRevLett.109.053002}
Matthieu Viteau, Paul Huillery, Mark~G. Bason, Nicola Malossi, Donatella
  Ciampini, Oliver Morsch, Ennio Arimondo, Daniel Comparat, and Pierre Pillet.
\newblock Cooperative excitation and many-body interactions in a cold {Rydberg}
  gas.
\newblock {\em Phys. Rev. Lett.}, 109:053002, Jul 2012.

\bibitem{PhysRevLett.113.023006}
N.~Malossi, M.~M. Valado, S.~Scotto, P.~Huillery, P.~Pillet, D.~Ciampini,
  E.~Arimondo, and O.~Morsch.
\newblock Full counting statistics and phase diagram of a dissipative {Rydberg}
  gas.
\newblock {\em Phys. Rev. Lett.}, 113:023006, Jul 2014.

\bibitem{NaturePhysics1183}
Ga\"etan Alpha, Yevhen Miroshnychenko, Tatjana Wilk, Amodsen Chotia, Matthieu
  Viteau, Daniel Comparat, Pierre Pillet, Antoine Browaeys, and Philippe
  Grangier.
\newblock Observation of collective excitation of two individual atoms in the
  {Rydberg} blockade regime.
\newblock {\em Nature Phys.}, 5:115--118, January 2009.

\bibitem{PhysRevLett.100.113003}
T.~A. Johnson, E.~Urban, T.~Henage, L.~Isenhower, D.~D. Yavuz, T.~G. Walker,
  and M.~Saffman.
\newblock {Rabi} oscillations between ground and {Rydberg} states with
  dipole-dipole atomic interactions.
\newblock {\em Phys. Rev. Lett.}, 100(11):113003, Mar 2008.

\bibitem{PhysRevLett.100.013002}
Ulrich Raitzsch, Vera Bendkowsky, Rolf Heidemann, Bj\"orn Butscher, Robert
  L\"ow, and Tilman Pfau.
\newblock Echo experiments in a strongly interacting {Rydberg} gas.
\newblock {\em Phys. Rev. Lett.}, 100(1):013002, Jan 2008.

\bibitem{youngeraithelnjp2008}
Kelly~Cooper Younge and Georg Raithel.
\newblock Rotary echo tests of coherence in {Rydberg}-atom excitation.
\newblock {\em New Journal of Physics}, 11(4):043006, 2009.

\bibitem{PhysRevX.2.031011}
J.~Nipper, J.~B. Balewski, A.~T. Krupp, S.~Hofferberth, R.~L\"ow, and T.~Pfau.
\newblock Atomic pair-state interferometer: Controlling and measuring an
  interaction-induced phase shift in {Rydberg}-atom pairs.
\newblock {\em Phys. Rev. X}, 2:031011, Aug 2012.

\bibitem{NaturePhysics10914}
S.~Ravets, H.~Labuhn, D.~Barredo, L.~Béguin, T.~Lahaye, and A.~Browaeys.
\newblock Coherent dipole-dipole coupling between two single {Rydberg} atoms at
  an electrically-tuned {F}\"orster resonance.
\newblock {\em Nature Phys.}, 10:914--917, Oct 2014.

\bibitem{PhysRevLett.112.183002}
D.~Barredo, S.~Ravets, H.~Labuhn, L.~B\'eguin, A.~Vernier, F.~Nogrette,
  T.~Lahaye, and A.~Browaeys.
\newblock Demonstration of a strong {Rydberg} blockade in three-atom systems
  with anisotropic interactions.
\newblock {\em Phys. Rev. Lett.}, 112:183002, May 2014.

\bibitem{Faoro:NatureComm10.1038}
R.~Faoro, B.~Pelle, A.~Zuliani, P.~Cheinet, E.~Arimondo, and P.~Pillet.
\newblock Borromean three-body {FRET} in frozen {Rydberg} gases.
\newblock {\em Nat. Commun.}, 6, Sep 2015.

\bibitem{PhysRevLett.115.013001}
R.~Celistrino Teixeira, C.~Hermann-Avigliano, T.~L. Nguyen,
  T.~Cantat-Moltrecht, J.~M. Raimond, S.~Haroche, S.~Gleyzes, and M.~Brune.
\newblock Microwaves probe dipole blockade and van der {W}aals forces in a cold
  {Rydberg} gas.
\newblock {\em Phys. Rev. Lett.}, 115:013001, Jun 2015.

\bibitem{Schau1455}
P.~Schau{\ss}, J.~Zeiher, T.~Fukuhara, S.~Hild, M.~Cheneau, T.~Macr{\`\i},
  T.~Pohl, I.~Bloch, and C.~Gross.
\newblock Crystallization in {I}sing quantum magnets.
\newblock {\em Science}, 347(6229):1455--1458, 2015.

\bibitem{PhysRevLett.114.203002}
A.~Urvoy, F.~Ripka, I.~Lesanovsky, D.~Booth, J.~P. Shaffer, T.~Pfau, and
  R.~L\"ow.
\newblock Strongly correlated growth of {Rydberg} aggregates in a vapor cell.
\newblock {\em Phys. Rev. Lett.}, 114:203002, May 2015.

\bibitem{PhysRevA.72.063403}
F.~Robicheaux and J.~V. Hern\'andez.
\newblock Many-body wave function in a dipole blockade configuration.
\newblock {\em Phys. Rev. A}, 72(6):063403, Dec 2005.

\bibitem{schausz2012}
Peter Schausz, Marc Cheneau, Manuel Endres, Takeshi Fukuhara, Sebastian Hild,
  Ahmed Omran, Thomas Pohl, Christian Gross, Stefan Kuhr, and Immanuel Bloch.
\newblock Observation of spatially ordered structures in a two-dimensional
  {Rydberg} gas.
\newblock {\em Nature}, 491(7422):87--91, 2012.

\bibitem{PhysRevA.87.053414}
David Petrosyan, Michael H\"oning, and Michael Fleischhauer.
\newblock Spatial correlations of {Rydberg} excitations in optically driven
  atomic ensembles.
\newblock {\em Phys. Rev. A}, 87:053414, May 2013.

\bibitem{PhysRevA.88.061406}
A.~Schwarzkopf, D.~A. Anderson, N.~Thaicharoen, and G.~Raithel.
\newblock Spatial correlations between {Rydberg} atoms in an optical dipole
  trap.
\newblock {\em Phys. Rev. A}, 88:061406, Dec 2013.

\bibitem{Gaj:NatureComm}
A.~Gaj, A.T. Krupp, J.B. Balewski, R.~Löw, S.~Hofferberth, and T.~Pfau.
\newblock From molecular spectra to a density shift in dense {Rydberg} gases.
\newblock {\em Nat. Commun.}, 5, Aug 2014.

\bibitem{Niederprum2016}
Thomas Niederpr{\"u}m, Oliver Thomas, Tanita Eichert, Carsten Lippe, Jes{\'u}s
  P{\'e}rez-R{\'i}os, Chris~H. Greene, and Herwig Ott.
\newblock Observation of pendular butterfly {Rydberg} molecules.
\newblock {\em Nat. Commun.}, 7, Oct 2016.

\bibitem{PhysRevA.70.042713}
Wenhui Li, Michael~W. Noel, Michael~P. Robinson, Paul~J. Tanner, Thomas~F.
  Gallagher, Daniel Comparat, Bruno Laburthe~Tolra, Nicolas Vanhaecke, Thibault
  Vogt, Nassim Zahzam, Pierre Pillet, and Duncan~A. Tate.
\newblock Evolution dynamics of a dense frozen {Rydberg} gas to plasma.
\newblock {\em Phys. Rev. A}, 70(4):042713, Oct 2004.

\bibitem{PhysRevLett.101.205005}
J.~P. Morrison, C.~J. Rennick, J.~S. Keller, and E.~R. Grant.
\newblock Evolution from a molecular {Rydberg} gas to an ultracold plasma in a
  seeded supersonic expansion of no.
\newblock {\em Phys. Rev. Lett.}, 101(20):205005, Nov 2008.

\bibitem{PhysRevLett.110.045004}
M.~Robert-de Saint-Vincent, C.~S. Hofmann, H.~Schempp, G.~G\"unter,
  S.~Whitlock, and M.~Weidem\"uller.
\newblock Spontaneous avalanche ionization of a strongly blockaded {Rydberg}
  gas.
\newblock {\em Phys. Rev. Lett.}, 110:045004, Jan 2013.

\bibitem{PhysRevLett.110.253003}
G.~Bannasch, T.~C. Killian, and T.~Pohl.
\newblock Strongly coupled plasmas via {Rydberg} blockade of cold atoms.
\newblock {\em Phys. Rev. Lett.}, 110:253003, Jun 2013.

\bibitem{PhysRevA.89.022701}
M.~Siercke, F.~E. Oon, A.~Mohan, Z.~W. Wang, M.~J. Lim, and R.~Dumke.
\newblock Density dependence of the ionization avalanche in ultracold {Rydberg}
  gases.
\newblock {\em Phys. Rev. A}, 89:022701, Feb 2014.

\bibitem{book:ramsey}
N.F. Ramsey.
\newblock {\em Molecular Beams}.
\newblock Oxford University Press, 1986.

\bibitem{PhysRevLett.104.133003}
Nicolas Saquet, Anne Cournol, J\'er\^ome Beugnon, Jacques Robert, Pierre
  Pillet, and Nicolas Vanhaecke.
\newblock {Landau}-{Zener} transitions in frozen pairs of {Rydberg} atoms.
\newblock {\em Phys. Rev. Lett.}, 104(13):133003, Apr 2010.

\bibitem{PhysRevA.59.988}
N.~V. Vitanov.
\newblock Transition times in the {Landau}-{Zener} model.
\newblock {\em Phys. Rev. A}, 59:988--994, Feb 1999.

\bibitem{landau1932}
L.D. Landau.
\newblock On the theory of transfer of energy at collisions ii.
\newblock {\em Phys. Z. Sowjetunion}, 2:46, 1932.

\bibitem{zener1932}
Clarence Zener.
\newblock Non-adiabatic crossing of energy levels.
\newblock {\em Proceedings of the Royal Society of London. Series A},
  137(833):696--702, 1932.

\bibitem{book:statisticaldistributions2000}
M.~Evans, N.~Hastings, and B.~Peacock.
\newblock {\em Statistical Distributions}.
\newblock Wiley-Interscience, 2000.

\bibitem{book:handbookInt2000}
Alan Jeffrey and Daniel Zwillinger.
\newblock {\em Table of Integrals, Series, and Products}.
\newblock San Diego, CA: Academic Press, 2000.

\bibitem{Amsler20081}
C.~Amsler \textit{et al.}
\newblock Review of particle physics.
\newblock {\em Physics Letters B}, 667(1-5):1 -- 6, 2008.

\bibitem{book:gelman}
A.~Gelman and J.~Hill.
\newblock {\em Data Analysis Using Regression and Multilevel/Hierarchical
  Models}.
\newblock Cambridge University Press, 2007.

\bibitem{book:PMTHandbook}
{\em Photomultiplier Handbook}.
\newblock Burle Technologies, inc., 1980.

\bibitem{Prescott1966173}
J.R. Prescott.
\newblock A statistical model for photomultiplier single-electron statistics.
\newblock {\em Nuclear Instruments and Methods}, 39:173 -- 179, 1966.

\end{thebibliography}

\end{document}